\documentclass[aps,prl,twocolumn,floats]{revtex4-2} 
\usepackage{amsmath}
\usepackage{graphicx}
\usepackage{dcolumn}
\usepackage{bm}
\usepackage{amssymb}
\usepackage{latexsym}
\usepackage{color}
\usepackage{subfigure}
\usepackage{ulem}

\begin{document}

\title{The Aharonov-Casher phase is geometrical and not topological}

\author{Igor Kuzmenko$^{1,2}$, Y. B. Band$^{1,2,3}$, Yshai Avishai$^{1,4}$}

\affiliation{
  $^1$Department of Physics,
  Ben-Gurion University of the Negev,
  Beer-Sheva 84105, Israel
  \\
  $^2$Department of Chemistry,
  Ben-Gurion University of the Negev,
  Beer-Sheva 84105, Israel
  \\
  $^3$The Ilse Katz Center for Nano-Science,
  Ben-Gurion University of the Negev,
  Beer-Sheva 84105, Israel
  \\
  $^4$Yukawa Institute for Theoretical Physics, Kyoto, Japan
  }

\begin{abstract}
It is demonstrated that the Aharonov-Casher (AC) phase is a geometric phase that, in general, depends on the details of the closed path taken by a particle with a magnetic moment that is subject to an electric field.  Consequently, it is not a topological phase. The proof of this statement is obtained by developing a counterexample that elucidates the dependence of the AC phase on the details of the path. Furthermore, we demonstrate that, in the particular example considered here, paths having an Abelian AC phase factor, also have an AC phase that is path-independent, whereas paths having a non-Abelian AC phase factor may have an AC phase that is path-dependent (i.e., not topological).
\end{abstract}

\maketitle

\begin{figure}
\centering
  \includegraphics[width=0.9\linewidth,angle=0] {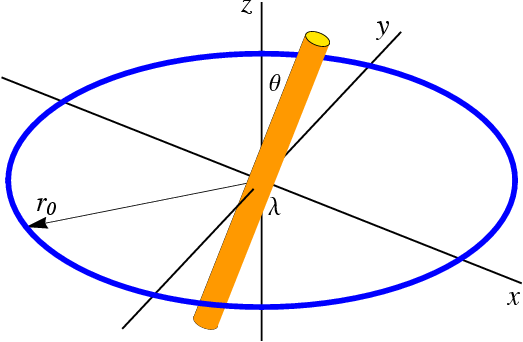}
\caption{\footnotesize Path $P$ (blue ring) encircles a line of charge with linear charge density $\lambda$.  The radius of the ring is $r_0$.  The line of charge is situated in the $x$-$z$ plane and makes an angle of $\theta$ with the $z$-axis.}
\label{Fig:Aharonov-Casher-circle}
\end{figure}

{\it Introduction}:
The Aharonov-Casher (AC) effect occurs when a particle with a magnetic moment moves along a closed path (e.g., a planar circular ring) in the presence of an electric field that threads the path and induces a spin-orbit coupling \cite{AC_84, Goldhaber_89, Boyer_87, Zeilinger_91, He_91, Reuter_91, Avishai_14, Cohen_19}, see Fig.~\ref{Fig:Aharonov-Casher-circle}.  It has been observed in a gravitational neutron interferometer \cite{Cimmino_89} and subsequently by fluxon interference of magnetic vortices in Josephson junctions \cite{Elion_93}, and it has also been measured for electrons \cite{Nagasawa_12, Nagasawa_13} and atoms \cite{Zeiske_95, Gillot_14}.  The AC phase of the particle's wavefunction is often designated a topological phase, meaning that the phase is not dependent on the details of the path taken by the particles, see for example Refs.~\cite{AC_84, Goldhaber_89, Zeilinger_91, He_91, Reuter_91, Dulat_12}.  As mentioned in Ref.~\cite{Reuter_91}, the AC phase is a special case of a Berry phase (see Ref.~\cite{AC_Berry} which comments on the relation between the two phases).

Here we prove, by presenting a counterexample, that the AC phase is not a topological phase, but rather a geometric phase that depends on the details of the path taken by the particle.  This is in contrast to the Aharonov-Bohm phase \cite{AB_59}, which occurs when the closed path of a charged particle is threaded by a magnetic flux.  The Aharonov-Bohm phase is a topological phase, meaning that it depends only on the magnetic flux threading the path and not on the details of the path itself.

Reference \cite{Goldhaber_89} focused on the case where the magnetic moment of the particle is oriented perpendicular to the plane of motion of the particle, and the electric field originates from an infinite line of charge aligned perpendicularly to the plane, so that ``the line charge must be straight and parallel to the magnetic moment''.   In this scenario, the AC phase factor is Abelian (see below). As will be demonstrated below, for an Abelian phase factor, the AC phase is independent of the details of the particle's path in the plane.  Moreover, if the path is circular, then the velocity (momentum) of the particle is normal to the electric field. Consequently, no force is exerted on the particle due to the line of charge (since the spin-orbit energy is constant as the particle moves in the circular orbit).  Reference~\cite{Boyer_87} addresses the relationship of the AC effect to the presence of classical forces, and asserts that: (a) ``a magnetic dipole particle passing a line charge does indeed experience a classical electromagnetic force'', that (b) the ``force will produce a relative lag between dipoles passing on opposite sides of the line charge'', that (c) ``the classical lag then leads to a quantum phase shift in exact agreement with that calculated by Aharonov and Casher'', and that (d) ``the proposed Aharonov-Casher effect has a transparent explanation as a classical lag effect'', and that the AC phase ``may actually involve a classical electromagnetic lag effect''.  Reference~\cite{SM} shows that when the particle's path is a circular ring, and the line of charge is perpendicular to the plane of motion, and  it intersects the center of the ring, no force is exerted on the particle in the direction of the ring. However, there is a non-vanishing AC phase, hence the AC effect is {\it not} related to the presence of a force.  Moreover, when the line of charge does not pass through the center of the circular ring, a non-vanishing force is exerted; nevertheless, the AC phase is identical to that obtained when the line of charge does pass through the center of the circular ring.

In Ref.~\cite{Peshkin_95},  the scalar Aharonov-Bohm effect was examined and the AC effect was also considered in passing.  The authors concluded that the AC effect is not topological.  However, Ref.~\cite{Dulat_12} asserts that Ref.~\cite{Peshkin_95} used the ``wrong Hamiltonian which yields their conclusion incorrect''.  In Ref.~\cite{SM} we consider this issue in detail using the Foldy-Wouthuysen transformation of the Dirac equation \cite{BD, Frohlich_93} to derive the correct Pauli Hamiltonian, and we demonstrate that, in general, a symmetrized form of the operators in the spin-orbit interaction term of the Hamiltonian is necessary.

The AC phase $\varphi_{\rm AC}$ is determined by the AC phase factor (in Gaussian units),
\begin{equation} \label{eq:U_AC}
{\mathcal U}_{\rm AC} = {\mathcal P} \exp \! \left[-\frac{i}{\hbar c} \oint {\boldsymbol \mu} \times {\bf E}({\bf r}) \cdot d{\bf r}\right] ,
\end{equation}
which is in general non-Abelian \cite{Avishai_14, Grosfeld_11, Wu_22, WuYang} [see Eq.~(\ref{eq:AC-phase-vs-Tr-U})]. Here ${\boldsymbol \mu}$ is the magnetic moment of the particle, ${\bf E}({\bf r})$ is the local electric field at position ${\bf r}$, and the path-ordered integration, indicated by ${\mathcal P}$, is over the closed path of the particle. Only under special circumstances is the phase factor Abelian (when the path lies in a plane and the electric field direction is in the same plane).  References \cite{AC_84, Goldhaber_89} explicitly consider this special case.  Otherwise, the AC phase factor is non-Abelian, and the AC phase depends on the specifics of the path and is therefore not topological.  In a more general context, when a magnetic moment is subject to an electric field, it is not guaranteed that an Abelian AC phase factor will yield a topological AC phase.  To illustrate this point, consider a circular ring lying in the $x$-$y$ plane and a constant electric field along the $z$-axis.  It is possible to apply a unitary transformation on the original Hamiltonian, thereby converting the AC phase factor to be Abelian.  Nonetheless, the AC phase is not topological because it depends on the radius of the ring, as demonstrated in Ref.~\cite{SM}.

{\it AC Hamiltonian}:
As a particular case of the AC effect, we consider a theoretical model system consisting of an uncharged spin-$1/2$ particle, such as a neutron or an alkali atom, in a waveguide that restricts the particle's movement to a planar ring.  The magnetic moment of the particle is $\boldsymbol\mu = g \mu_B {\boldsymbol \sigma}/2$, where $\boldsymbol\sigma = (\sigma_x, \sigma_y, \sigma_z)$ is the Pauli spin matrix.  The ring is threaded by a line of charge with linear charge density $\lambda$, that is situated in the $x$-$z$ plane and makes an angle of $\theta$ with the $z$-axis, as illustrated in Fig.~\ref{Fig:Aharonov-Casher-circle}.
Hereafter, we use cylindrical coordinates ${\bf r} = (r, \phi, z)$, wherein the path on the ring of radius $r_0$ in the $x$-$y$ plane, centered at the origin of the coordinate system, is given by the vector-valued function ${\bf r} (\phi) = r_0 \, (\cos \phi \, \hat{\bf x} + \sin \phi \, \hat{\bf y}) = r_0 {\bf e}_{r}$.

The line of charge generates an electric field ${\bf E} ({\bf r} (\phi), \theta) = \tfrac{\lambda}{r_0} \boldsymbol{\mathcal E} (\phi, \theta)$ at position ${\bf r} = {\bf r} (\phi)$ on the path, where the dimensionless electric field $\boldsymbol{\mathcal E} (\phi, \theta)$ is given by
\begin{eqnarray}
  \boldsymbol{\mathcal E} (\phi, \theta) &=&
  2 \, {\bf e}_{r} +
  \frac{\sin^2 \theta \, \sin (2 \phi)}{\cos^2 \theta \, \cos^2 \phi + \sin^2 \phi} \, {\bf e}_{\phi}
  \nonumber \\ && -
  \frac{\sin (2 \theta) \, \cos \phi}{\cos^2 \theta \, \cos^2 \phi + \sin^2 \phi} \, {\bf e}_{z} .
  \label{eq:E-line}
\end{eqnarray}

The particle  with mass $M$, velocity ${\bf v}$, and momentum ${\bf p} = M {\bf v}$, moving along the path, is subject to an effective magnetic field ${\bf B}({\bf r} (\phi), \theta) = {\bf E} ({\bf r} (\phi), \theta) \times {\bf v}/c$ in its rest frame.  
The momentum operator of a particle moving along path $P$ can be expressed as
${\bf p} = {\bf e}_{\phi} \, p$, where
\begin{equation}
  p = - \frac{i \hbar}{r_0} \, \frac{d}{d \phi} .
\end{equation}
The Pauli Hamiltonian,
\begin{eqnarray}   \label{eq:H_AC}
H (\theta) &=&
\frac{1}{2 M} \,
\Big( p + \frac{\hbar}{r_0} \, {\boldsymbol \sigma} \cdot \boldsymbol {\mathcal{A}} (\phi, \theta) \Big)^{2}
+ \frac{\lambda^2 \alpha}{2 r_0^2} \boldsymbol{\mathcal E}^2 (\phi, \theta)
\nonumber \\ && + \frac{\hbar^2}{2 M r_0^2} \, \frac{\lambda^2}{\lambda_0^2} \,
\Big( 3 \, \mathcal{E}^{2} (\phi, \theta) + \mathcal{E}_{\phi}^{2} (\phi, \theta) \Big) ,
\end{eqnarray}
for the particle is obtained by making the Foldy-Wouthuysen transformation \cite{SM, BD, Frohlich_93, Band-Avishai-QuantumMechanics}, where $\alpha$ is the particle polarizability, $\mathcal{E}_{r} (\phi, \theta)$, $\mathcal{E}_{\phi} (\phi, \theta)$ and $\mathcal{E}_{z} (\phi, \theta)$ are cylindrical components of $\boldsymbol{\mathcal E} (\phi, \theta)$ in Eq.~(\ref{eq:E-line}),
$\mathcal{E} (\phi, \theta) = |\boldsymbol{\mathcal E} (\phi, \theta)|$, and
the dimensionless vector potential $\boldsymbol {\mathcal{A}} (\phi, \theta)$ is
\begin{eqnarray}
  {\boldsymbol {\mathcal{A}}} (\phi, \theta) &=&
  \frac{g \mu_B \lambda}{4 \hbar c} \, \boldsymbol{\mathcal E} (\phi, \theta)  \times{\bf e}_{\phi} =
  \frac{\lambda}{\lambda_0}
  \bigg[
    {\bf e}_{z}
    \nonumber \\ && +
    \frac{1}{2} \,
    \frac{\sin (2 \theta) \, \cos (\phi)}{\cos^2 (\theta) \, \cos^2 (\phi) + \sin^2 (\phi)} \,
    {\bf e}_{r}
  \bigg] .
  \label{eq:vector-potential-SU2}
\end{eqnarray}
Note that in the symmetric form of the spin-orbit interaction, the two operators that are linear in $p$ are generically not equal (see Ref.~\cite{SM}).  The parameter $\lambda_0$ has units of the linear charge density and is given by $\lambda_0 = \frac{2 \hbar c}{g \mu_B}$.  Note that the dimensionless vector potential $\boldsymbol{\mathcal{A}} (\phi, \theta)$ is independent of the path radius $r_0$, but both the magnitude and the direction of the vector $\boldsymbol{\mathcal{A}} (\phi, \theta)$ depend on the angle $\theta$.  We mention in passing that if the charge density of the line of charge is time-dependent, $\lambda(t)$, then the electric field, $\boldsymbol{\mathcal E} ({\bf r}, t)$, and the vector potential, ${\boldsymbol {\mathcal{A}}}({\bf r}, t)$, are time-dependent; consequently, the Hamiltonian in Eq.~(\ref{eq:H_AC}) is time-dependent.  This results in a dynamical AC effect that can be treated with the time-dependent Schr\"odinger equation or, if the rate of change of the charge density were slow, with the adiabatic approximation.

{\it Bound state wave functions and energies}:
The wave functions $\Psi (\phi, \theta)$ and the energies $\epsilon (\theta)$ of particles that move along path $P$ are found from the Schr\"odinger equation
\begin{equation}   \label{eq:Schrodinger-AC}
  H (\theta) \, \Psi (\phi, \theta) = \epsilon (\theta) \, \Psi (\phi, \theta) ,
\end{equation}
where the Hamiltonian $H (\theta)$ is given in Eq.~(\ref{eq:H_AC}).  The wave function satisfies the periodic boundary condition
\begin{equation}   \label{eq:Psi-periodic}
  \Psi (\phi+ 2 \pi, \theta) = \Psi (\phi, \theta) .
\end{equation}

It is convenient to apply a gauge transformation which simplifies the Hamiltonian $H (\theta)$,
\begin{equation}   \label{eq:gauge-transform}
  \Psi (\phi, \theta) = \mathcal{U} (\phi, \theta) \tilde\Psi (\phi, \theta) ,
\end{equation}
where the AC phase factor $\mathcal{U} (\phi, \theta)$ is a unitary matrix defined as a path-ordered integral:
\begin{eqnarray}    \label{eq:U-def}
  \mathcal{U} (\phi, \theta) &=&
  {\mathcal P}\!\exp \!
  \bigg[
    - i \int\limits_{0}^{\phi}
    \boldsymbol\sigma \cdot \boldsymbol{\mathcal{A}} (\phi', \theta) \, d \phi'
  \bigg]
  \nonumber \\ &\equiv&
  \sum_{n = 0}^{\infty}
  \big( - i \big)^{n}
  \prod_{l = 1}^{n}
  \int\limits_{0}^{\phi_{l - 1}}
  \boldsymbol\sigma \cdot \boldsymbol{\mathcal{A}} (\phi_l, \theta) \, d \phi_l ,
\end{eqnarray}
where, in the upper limit of integration of the second equation of Eq.~(\ref{eq:U-def}), $\phi_0 = \phi$.
The transformed wave function $\tilde\Psi (\phi, \theta)$ satisfies the simple Schr\"odinger equation
\begin{equation}   \label{eq:Schtodinger-tilde}
  \tilde{H} (\theta) \, \tilde\Psi (\phi, \theta) = \epsilon (\theta) \, \tilde\Psi (\phi, \theta) ,
\end{equation}
where
\begin{eqnarray}
  \tilde{H} (\theta) &\equiv&
  \mathcal{U}^{-1} (\phi, \theta) H (\theta)  \mathcal{U} (\phi, \theta) = \frac{p^2}{2 M} 
  + \frac{\lambda^2 \alpha}{2 r_0^2} \boldsymbol{\mathcal E}^2 (\phi, \theta)
  \nonumber \\ &&
  + \frac{\hbar^2}{2 M r_0^2} \, \frac{\lambda^2}{\lambda_0^2} \,
  \Big( 3 \, \mathcal{E}^{2} (\phi, \theta) + \mathcal{E}_{\phi}^{2} (\phi, \theta) \Big) .
  \label{eq:H-tilde}
\end{eqnarray}
Note that the Hamiltonian $\tilde{H} (\theta)$ contains no spin-orbit interaction; the spin-orbit interaction has been transformed away.  Instead, the boundary condition on $\tilde\Psi (\phi, \theta)$, which depends on the angle $\theta$ between the line of charge and the $z$-axis, will incorporate the effect of the spin-orbit interaction.  In place of the periodic boundary condition on the wave function $ \Psi (\phi, \theta)$ in Eq.~(\ref{eq:Psi-periodic}), the boundary condition on $\tilde\Psi (\phi, \theta)$,
\begin{equation}   \label{eq:Psi-periodic-tilde}
  \tilde\Psi (\phi + 2 \pi, \theta) = \mathcal{U}^{-1} (2 \pi, \theta) \, \tilde\Psi (\phi, \theta) ,
\end{equation}
is required to ensure the correct transformation properties of the wave function.  The SU(2) unitary matrix ${\mathcal U}_{\rm AC} (\theta) \equiv \mathcal{U} (2 \pi, \theta)$ [see Eq.~(\ref{eq:U-def})] can be represented as
\begin{equation}   \label{eq:U-vs-phi_AC}
  \mathcal{U}_{\rm AC} (\theta) =
  e^{- i \varphi_{\rm AC}(\theta) \, \hat{\bf b}(\theta) \cdot \boldsymbol\sigma} ,
\end{equation}
where $\varphi_{\rm AC}(\theta)$ is a real parameter called the AC phase, and $\hat{\bf b}(\theta)$ is a real unit vector.  The AC phase $\varphi_{\rm AC}(\theta)$ and the unit vector $\hat{\bf b}(\theta)$ can be determined in terms of $\mathcal{U}_{\rm AC} (\theta)$ as follows: 
\begin{eqnarray}
  \cos \varphi_{\rm AC}(\theta) &=&
  \frac{1}{2} \, {\rm Tr} \big[ \mathcal{U}_{\rm AC} (\theta) \big] ,   \label{eq:cos_AC_phase}
  \\
  \sin \varphi_{\rm AC}(\theta) ~ \hat{\bf b}(\theta) &=&
  \frac{i}{2} \, {\rm Tr} \big[ \mathcal{U}_{\rm AC} (\theta) \boldsymbol\sigma \big] . \label{eq:unit_b}
\end{eqnarray}
Both $\varphi_{\rm AC}(\theta)$ and $\hat{\bf b}(\theta)$ are dependent on the linear charge density $\lambda$ and the angle $\theta$, and are independent of $r_0$.  When $\theta = 0$, $\hat{\bf b}(0) = \hat{\bf z}$, and $\varphi_{\rm AC}(0) = 2 \pi \, \lambda / \lambda_0$ [see the discussion below Eq.~(\ref{eq:dU=Phi-U})].

{\it Calculation of the Aharonov-Casher phase}:
In order to calculate the AC phase, we employ the differential
equation for $\mathcal{U} (\phi, \theta)$:
\begin{equation}   \label{eq:dU=sigma-B-U}
  \frac{d \mathcal{U} (\phi, \theta)}{d \phi} =
  - i \,
  \boldsymbol\sigma \cdot \boldsymbol{\mathcal{A}} (\phi, \theta) \,
  \mathcal{U} (\phi, \theta) .
\end{equation}
As $\phi$ approaches zero, the matrix $\mathcal{U} (\phi, \theta)$ converges to the 2$\times$2 dimensional identity matrix.  The derivation of Eq.~(\ref{eq:dU=sigma-B-U}) is accomplished by differentiating Eq.~(\ref{eq:U-def}) with respect to $\phi$.  Once the matrix $\mathcal{U}_{\rm AC} (\theta)$ is determined, the AC phase can be calculated using the equation,
\begin{equation}   \label{eq:AC-phase-vs-Tr-U}
  \varphi_{\rm AC}(\theta) = \arccos  \Big( \frac{1}{2} \, {\rm Tr} \big[ \mathcal{U}_{\rm AC} (\theta) \big] \Big) .
\end{equation}
where the $\arccos$ function needs to be properly analytically continued (see below).  Notice that the AC phase $\varphi_{\rm AC}(\theta)$ does not depend on the ring radius $r_0$ but it does depend on $\theta$.

\begin{figure}
\centering
  \includegraphics[width=0.9\linewidth,angle=0] {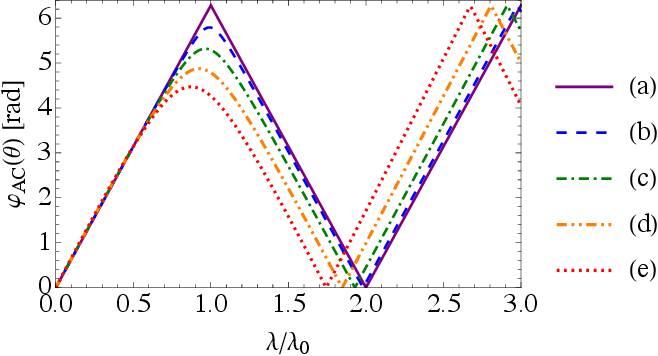}
\caption{\footnotesize AC phase $\varphi_{\rm AC}(\theta)$ (in radians) versus the dimensionless charge density variable $\lambda / \lambda_0$ for a variety of values of the angle $\theta$: (a) $\theta = 0$, (b) $\theta = \pi / 20$, (c) $\theta = \pi / 10$, (d) $\theta = 3 \pi / 20$ and (e) $\theta = \pi / 5$.
}
\label{Fig:AC-phase}
\end{figure}

Figure~\ref{Fig:AC-phase} depicts the AC phase $\varphi_{\rm AC}(\theta)$ versus the dimensionless charge density variable $\lambda / \lambda_0$, which has been numerically calculated, for a variety of values of the angle $\theta$.  Note that $\varphi_{\rm AC}(\theta)$ depends on the angle $\theta$ (since the vector potential does), hence it is {\it not a topological phase}, but rather {\it a geometrical phase}, because changing the plane of the ring effectively means changing $\theta$.  Moreover, the AC phase factor is non-Abelian for $\theta \ne 0$, see Eq.~(\ref{eq:U-def}).  For $\theta = 0$, the AC phase factor is Abelian, and Eq.~(\ref{eq:dU=sigma-B-U}) takes the form
\begin{equation}   \label{eq:dU=Phi-U}
  \frac{d \mathcal{U} (\phi, 0)}{d \phi} =
  - i \, \frac{\lambda \sigma_z}{\lambda_0} \, \mathcal{U} (\phi, 0) .
\end{equation}
The AC phase for $\theta = 0$ obtained from this equation is $\varphi_{\rm AC}(0) = 2 \pi \lambda / \lambda_0$, hence $\cos \varphi_{\rm AC}(0)$ is a periodic function of $\lambda$ with period $\lambda_0$.  The AC phase $\varphi_{\rm AC}(0)$ in Eq.~(\ref{eq:AC-phase-vs-Tr-U}) is a multivalued function of $\lambda$.  Therefore, if $\varphi_{\rm AC}(0)$ is a solution of Eq.~(\ref{eq:cos_AC_phase}) for $\theta = 0$, then $\varphi_{\rm AC}(0) + 2 \pi n$ and $-\varphi_{\rm AC}(0) + 2 \pi n$, where $n$ is integer, are also solutions.  The solid purple curve showing the AC phase for $\theta = 0$ in Fig.~\ref{Fig:AC-phase} is a plot of the piecewise function
\begin{equation}   \label{eq:AC-phase-piecewise}
  \varphi_{\rm AC}(0) =
  \left\{
    \begin{array}{ccc}
      \displaystyle
      2 \pi \Big( \frac{\lambda}{\lambda_0} - 2 n \Big),
      &\text{for}&
      \displaystyle
      2 n \leq \frac{\lambda}{\lambda_0} < 2 n + 1 ,
      \\
      \displaystyle
      2 \pi \Big( 2 n - \frac{\lambda}{\lambda_0} \Big) ,
      &\text{for}&
      \displaystyle
      2 n - 1 \leq \frac{\lambda}{\lambda_0} < 2 n .
    \end{array}
  \right.
\end{equation}
$\varphi_{\rm AC}(0)$ is a periodic function of $\lambda$ with period $2 \lambda_0$.  When $\theta \neq 0$, the AC phase $\varphi_{\rm AC}(\theta)$ is not periodic in $\lambda$; it increases with $\lambda$ and approaches $2 \pi$ at $\lambda$ close to $\lambda_0$ and then decreases, reaches a positive minimum, and then increases (the oscillations of $\varphi_{\rm AC}(\theta)$ with increasing $\lambda$ are shown in Fig.~\ref{Fig:AC-phase}).  For large $\lambda$, $\varphi_{\rm AC}(\theta)$ is nearly a linear piecewise function of $\lambda$ and  varies from $0$ to $2 \pi$ over a large range of $\lambda$.

\begin{figure}
\centering
  \includegraphics[width=0.9\linewidth,angle=0] {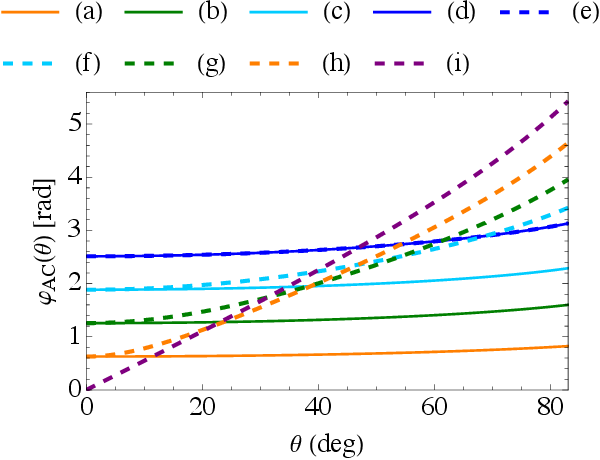}
\caption{\footnotesize AC phase $\varphi_{\rm AC}(\theta)$ (in radians) as a function of the angle $\theta$ for
a variety of values of the linear charge density $\lambda$:
(a) $\lambda = 0.1 \lambda_0$, (b) $\lambda = 0.2 \lambda_0$,
(c) $\lambda = 0.3 \lambda_0$, (d) $\lambda = 0.4 \lambda_0$,
(e) $\lambda = 0.6 \lambda_0$, (f) $\lambda = 0.7 \lambda_0$,
(g) $\lambda = 0.8 \lambda_0$, (h) $\lambda = 0.9 \lambda_0$
and (i) $\lambda = \lambda_0$. 
}
\label{Fig:AC-phase-theta}
\end{figure}

Figure~\ref{Fig:AC-phase-theta} plots the AC phase $\varphi_{\rm AC}(\theta)$ versus the angle $\theta$ for a variety of values of the linear charge density $\lambda$.  When $\theta = 0$, $\varphi_{\rm AC}(0)$ is given by Eq.~(\ref{eq:AC-phase-piecewise}); for linear charge densities within the range $0 \leq \lambda \leq \lambda_0$, the AC phase is symmetric with respect to the transformation $\lambda \to \lambda_0 - \lambda$, so $\varphi_{\rm AC}(0) = 0$ for $\lambda = 0$ and $\lambda_0$, and $\varphi_{\rm AC}(0) = \pi$ at $\lambda = \lambda_0/2$.  In contrast, when $\theta$ is non-zero, this symmetry is lifted, and $\varphi_{\rm AC}(\theta)$ is not linear in $\lambda$.  Clearly, the AC phase depends on $\theta$ and is therefore {\it not topological}.

{\it Summary and Conclusions}:
We proved by counterexample that the AC phase is not a topological phase; it depends on the details of the path that is threaded by the line of charge.  In this sense, the AC phase is different from the Aharonov-Bohm phase, which is topological.  However, as noted in the text after Eq.~(\ref{eq:U_AC}), the non-topological nature of the AC phase is due to the non-Abelian nature of the AC phase factor.
In the special case where the path lies in a plane, say the $x$-$y$ plane, and the direction of the electric field is in this same plane (i.e., the line of charge is normal to the $x$-$y$ plane), then the AC phase factor ${\mathcal U}_{\rm AC}$ is Abelian, and the AC phase is independent of the details of the path threaded by the line of charge. In this special case the direction of the effective magnetic field acting on the particle in its rest frame is constant along the closed path.  Quite generally, the AC effect is {\it not} related to the presence of a force \cite{SM}.

It is important to note that the lack of topological character of the AC effect precludes the possibility of topological protection against local perturbations; this can adversely affect fault tolerance.  Quantum computing based on the AC effect has been discussed in Refs.~\cite{Ericsson_02, Ionicioiu_03, Bakke_11, Melo_14} and references therein.  It should be noted that the use of the AC effect in quantum computing or quantum information applications may therefore be affected by fault tolerance issues.

We thank Professor Yuval Gefen and Professor Lev Weidman for useful comments. 


\end{document}